\documentclass[12pt]{iopart}
\usepackage{iopams}
\usepackage{amsfonts}
\usepackage{amssymb}
\usepackage{dcolumn}   % needed for some tables
\usepackage{graphicx}
\usepackage{float}
\usepackage{mathrsfs}
\usepackage{amsbsy}
\usepackage{standalone}
\usepackage{filehook}
\usepackage{gincltex}
\usepackage{collectbox}
\usepackage{filemod-expmin}
\usepackage{cancel}
\usepackage{url}
\usepackage{tikz}

\newcommand{\text}[1]{\textrm{#1}}

\usepackage{calrsfs}
\DeclareMathAlphabet{\pazocal}{OMS}{zplm}{m}{n}

\makeatletter
\newcommand*\bigcdot{\mathpalette\bigcdot@{.5}}
\newcommand*\bigcdot@[2]{\mathbin{\vcenter{\hbox{\scalebox{#2}{$\m@th#1\bullet$}}}}}
\makeatother

\newtheorem{definition}{Definition}[section]

\begin{document}

Pre-print version - Please cite: \textcolor{blue}{https://doi.org/10.1088/1361-6382/ad0422}

\markboth{Robert Monjo}
{Galaxy rotation curve in hyperconical universes: a natural relativistic MOND}

%%%%%%%%%%%%%%%%%%%%%%%%%%%%%%%%%%%%%%%%%%%%%%%%%%%%%%%%%%%%%%%%%%%%% Universal Relativity: Cosmological print in GR

\title[Galaxy rotation curve in hyperconical universes: a natural relativistic MOND]{Galaxy rotation curve in hyperconical universes: a natural relativistic MOND}

 %\footnote{For the title, try not to use more than 3 lines.
%Typeset the title in 10~pt Times roman, uppercase and boldface.}  }

\author{Robert Monjo} % \footnote{Typeset names in
%~pt roman, uppercase. Use the footnote to indicate the
%present or permanent address of the author.}}

\address{Department of Algebra, Geometry and Topology, Complutense University of Madrid,\\ Plaza de las Ciencias 3, E-28040 Madrid, Spain, rmonjo@ucm.es}

\begin{abstract}
Modified Newtonian dynamics (MOND) and similar proposals can (at least partially) explain the excess rotation of galaxies or the equivalent mass-discrepancy acceleration, without (or by reducing) the requirement of dark matter halos. This paper develops a modified gravity model to obtain local limit to the general relativity (GR) compatible with a cosmological metric different to the standard Friedmann–Lemaître–Robertson–Walker metric. Specifically, the paper uses a distorted stereographic projection of hyperconical universes, which are 4D hypersurfaces embedded into 5D Minkowski spacetime. This embedding is a key in the MOND effects found in galactic scales. To adequately describe the mass-discrepancy acceleration relation, centrifugal force would present a small time-like contribution at large-scale dynamics due to curvature of the Universe. Therefore, the Lagrangian density is very similar to the GR but with subtracting the background curvature (or vacuum energy density) of the perturbed hyperconical metric. Results showed that the proposed model adjusts well to 123 galaxy rotation curves obtained from the Spitzer Photometry and Accurate Rotation Curves database, using only a free parameter.
% (1) local curvature of the Universe is perturbed by the baryonic matter density
\end{abstract}


%\noindent PACS numbers:xxx

%\tableofcontents
%Cosmology: Have we returned to the starting point?
\begin{table}[h]
    \centering
    \resizebox{9.91cm}{!}{
    \begin{tabular}{c|c}
       ADM &  Arnowitt-Deser-Misner \\
       CDM & cold dark matter \\
       BTFR & baryonic Tully–Fisher relation \\
       CMB  & cosmic microwave background \\
       EFE  &  external field effects \\ FLRW &
       Friedmann–Lemaître–Robertson–Walker \\
       GR & general relativity \\
       $\Lambda$CDM & dark energy and cold dark matter \\
       MDAR & mass-discrepancy acceleration relation \\
       MOND & modified Newtonian dynamics \\
       PPN & parameterized post-Newtonian\\
       RMAE & relative mean absolute error \\
       SEP & strong equivalence principle \\
       SPARC & Spitzer Photometry and Accurate Rotation Curves 
     \\
    TRP & thermal radiation pressure\\
      TRF & thermal recoil force \\
       WEP & weak equivalence principle
    \end{tabular}}
    \caption{List of abbreviations and acronyms used in this paper.}
    \label{tab:table1}
\end{table}

%3750 > 3478 + 205 (150/110*150) = 3683
\section{Introduction}
\label{sec:introduction}

\subsection{Limits of general relativity?}

General relativity (GR; see Table \ref{tab:table1} for a list of abbreviations and acronyms) has been validated at least for local scales, such as the solar system, among others
\cite{Ciufolini2019, Touboul2022}.  To achieve this, parameterized post-Newtonian (PPN) techniques and the weak equivalence principle (WEP, equalizing the inertial and gravitational masses) are usually considered \cite{Dittus2007, Liu2022}. For instance, employing the Pantheon sample of Type Ia supernova data, Liu et al. \cite{Liu2022} showed that GR is better fitted to the observations than the PPN model. Similar successes are achieved by the GR-based standard cosmological model ($\Lambda$ cold dark matter [CDM]) fitted to cosmic microwave background (CMB) data, for instance, when it models the first peak location of baryon acoustic oscillations \cite{Aghanim2020}.

Nevertheless, the strong equivalence principle (SEP) is recognized as the most appropriate to distinguish GR from other viable theories of gravity \cite{Chae2020}.  The SEP affirms that internal dynamics of a self-gravitating system under free fall (into an external gravitational field) should not depend on the external field strength.  In other words, every observer falling into a gravitational field can choose a locally inertial coordinate system for a sufficiently small region such that the laws of nature take the same form as in an unaccelerated Minkowskian frame in absence of gravitation \cite{Chae2020, Goto2010}.  However, according to the results of Chae et al. \cite{Chae2020}, external field effects (EFE) are statistically detected at more than 4$\sigma$ from 153 rotating galaxies of the Spitzer Photometry and Accurate Rotation Curves (SPARC) database. This outcome points to a breakdown of the SEP, supporting modified gravity theories beyond GR, since tidal effects from neighboring galaxies in the standard $\Lambda$CDM framework are not enough to explain it. 

Applying the standard gravity model, excess rotation appears in most galaxies, highlighting a discrepancy between the visible matter and the required mass to support the observed velocities. To solve this problem, galaxy rotation curves are usually modeled by the hypothetical presence of cold dark matter (CDM) in the halos \cite{Bhattacharya2013, Beck2023}. However, these hypothetical halos predict a systematically deviating relation from the observations, while all aspects of rotation curves appear to be more naturally explained by modified gravities \cite{McGaugh2007, Chae2022}. Furthermore, the empirically derived densities of the dark-matter halos for low-mass galaxies are half of what is predicted by CDM simulations \cite{deBlok2008}. %%%%–

Alternatively, modified gravities could explain the empirical mass-discrepancy acceleration relation, since it is found as a function of visible matter $M_b\,$ \cite{Trippe2014, Merritt2017, Goddy2023}. In other words, the local ratio between observed and expected velocities is strictly predictable given only the observed distribution of visible matter in the galaxy, without the need of additional (e.g. dark-matter-based) variables. The empirical law is known as mass-luminosity relation or baryonic Tully-Fisher relation (BTFR),\begin{equation} \label{eq:BTFR0}
M_b =  A\,v^\epsilon
\end{equation} for a `flat' velocity curve $v$ observed in a galaxy disk, with $\epsilon \approx 4$ (see for instance \cite{Goddy2023}). Or similarly, it can be expressed as a mass-discrepancy acceleration relation (MDAR) \cite{McGaugh2004,DiCintio2016},
\begin{eqnarray} \nonumber
\frac{M_{b}+M_{CDM}}{M_b} & =  \frac{v^2}{ v_K^2} = C |a_N|^{-\beta} = C\left(\frac{r}{v_K^2}\right)^\beta\;\; \Longrightarrow\;\; 
\\ \label{eq:discrepancyMDAR}
&\;\; \Longrightarrow\;\; 
v^{4} = C^2 r^{2\beta}\,v_K^{4(1-\beta)} \approx C^2 \mathrm{G}M_b \;\;
\end{eqnarray}where $\beta \approx 0.5$, $M_b+M_{CDM}$ is the total mass including CDM and $C^2\mathrm{G} \approx: A^{-1}$ is approximately inverse of the same constant as used in Eq. \ref{eq:BTFR0}, while Newtonian acceleration $|a_N| = \mathrm{G}M_b/r = v_K^2/ r$ is expressed as a function of the Kepler speed $v_K$ and the distance $r$ to the galactic center.  To solve the raised problem, some authors suggested that dark matter presents a stronger coupling to the baryon matter and, therefore, both matter contents are linked by an effective law \cite{Blanchet2007, Katz2016, Barkana2018}. However, the excess rotation only occurs where the acceleration induced by the visible matter is lower than a typical scale (almost constant value) of gravitation, pointing to a more general problem involving spatial scales than a problem involving matter types, as the relic galaxies seem to point \cite{Comeron2023}. Therefore, most of observations suggest the need to modify the standard gravity models \cite{Trippe2014, Merritt2017}.

\subsection{Limits of Newtonian dynamics?}

In 1983, Milgrom \cite{Milgrom1983} proposed modified Newtonian dynamics (MOND) as a possible alternative to the cold dark mass hypothesis. Milgrom's law is expressed in terms of the external force $F$ and the acceleration $a$ experienced by the objects,\begin{equation}
    F = m\ a\ \cdot \mu\left(\frac{a}{a_0}\right)\,,
\end{equation}where $a_0$ was proposed to be a new universal constant, valued about $a_0 \approx 1.2 \cdot 10^{-10}$ m/s${}^2$. That is the second Newton's law but multiplied by a scalar function $\mu:\mathbb{R}_{\ge 0} \to \mathbb{R}_{\ge 0}$ such that $\mu\left(\frac{a}{a_0}\right) \approx 1$ for $a>>a_0$, recovering Newton's $F = ma$, but it presents a `deep MOND behaviour' for $a \ll a_0$,\begin{equation}
F = m\frac{a^2}{a_0}\,,
\end{equation} or an equivalent in a Newtonian gravity system of mass $M=M_b$,\begin{equation}
\frac{\mathrm{G}Mm}{r^2} = m\frac{\left(\frac{v^2}{r}\right)^2}{a_0}\;\;\Longrightarrow\;\;v^4 = \mathrm{G}Ma_0\;,
\end{equation}which corresponds to the typical `flat velocity' of most of galaxy rotation curves. A remarkable finding on the value of $a_0$ is its possible relation with the age $t$ of the universe, since $a_0 \approx \frac{1}{6}c/t$ \cite{Milgrom1994, Blanchet2007}. Therefore, the current study explores a connection to the cosmological models.

MOND and other similar approaches can (at least partially) explain the excess rotation of galaxies or the equivalent mass-discrepancy acceleration without the requirement of dark-matter halos \cite{Gentile2008,Asencio2022}.  According to the results of \cite{Asencio2022}, the gravitational distortion of dwarf galaxies supports solutions based on modified gravity theories rather than the existence of dark matter. In particular, the lack of low surface brightness in dwarfs towards its center is incompatible with $\Lambda$CDM expectations but consistent with MOND.

Alternative works on modified gravity are based on \textit{Verlinde’s emergent gravity} \cite{Verlinde2011, Yoon2023}. For instance, Yoon et al. \cite{Yoon2023} claim that anomalous acceleration in galaxies is given by the Verlinde’s parameter that comes from cosmological observations of the expansion of the universe, independently of the rotation curve measurements. The Verlinde's model proposes a total gravitational acceleration $\mathrm{a}_{Ver}$ such as $\mathrm{a}_{Ver} = \sqrt{\mathrm{a}_{N}^2 + \mathrm{a}_{D}^2}$, where $\mathrm{a}_{N}$ is Newtonian gravity and $\mathrm{a}_{D}^2 \propto \frac{1}{6}cH_0 := a_0$ is the ``apparent dark-matter gravity'', which is hypothesised to be connected to the current Hubble parameter $H_0$. The anomalous acceleration that emerged from this model can be assimilated as a deep MOND regime.
 
However, MOND-like theories present important limitations to explain some observed dynamics, such as those found in the Ly$\alpha$ Forest and the Bullet Cluster \cite{Aguirre2001, Clowe2006, Angus2006}. This issue could be solved by a more general theory with variable $a_0$ and/or with a relativistic formulation with MOND behavior at the limit $a \ll a_0$.

\subsection{Relativistic formulation of MOND?}

%https://iopscience.iop.org/article/10.1088/0004-6256/136/6/2648#aj266574app1
%https://www.arxiv-vanity.com/papers/astro-ph/0509519/

Relativistic formulations of gravity with MOND behavior was initially proposed by Bekenstein and Milgrom in 1984 \cite{Bekenstein1984} but presented problems of superluminical solutions that were solved later by modifying the `physical metric' to $\tilde g_{\alpha\beta} = e^{-2\phi} g_{\alpha\beta} + u_\alpha u_\beta\,\left(e^{-2\phi}-e^{2\phi}\right)$ with vector field $u_\alpha$ dominated by the Lagrangian  \begin{equation}
{\mathcal {L}}_{v}=-{\frac {1}{16\pi \mathrm{G}}}\left[\left(\frac{K}{2}\,F^{\alpha \mu }F_{\alpha \mu }\right)+4 \lambda \left(g^{\mu \nu }u_{\mu }u_{\nu }-1\right)\right]{\sqrt {-g}}\end{equation}where $\mathrm{G}$ is the Newtonian constant (with $c \equiv 1$), $K$ is a new constant, $F_{\alpha\beta} = \partial_{\alpha}u_{\beta}-\partial_{\beta }u_{\alpha}$ and $\lambda$ is a Lagrange multiplier field included to enforce the normalization of the vector field, while the scalar function $\phi$ depends on the dynamics of the tensor $\mathrm{h}_{\alpha\beta} := g_{\alpha\beta} + u_\alpha u_\beta$. With these ingredients, Bekenstein defined the Tensor–vector–scalar gravity (TeVeS) model in 2004 \cite{Bekenstein2004}. However, TeVeS shows some problems in simulating stars, which are highly unstable on the scale of approximately two weeks \cite{Seifert2007}. Therefore, the construction of TeVeS requires an undetermined number of terms to solve all the issues raised \cite{Mavromatos2009}. Furthermore, any covariant theory of MOND should also be valid on cosmological scales \cite{Famaey2012}, since GR and the Friedmann–Lemaître–Robertson–Walker (FLRW) metric are the basis of the Friedmann equations (used in the standard $\Lambda$CDM model).

In this perspective, Skordis and Złośnik proposed an alternative relativistic MOND with additional terms analogous to the FLRW action \cite{Skordis2021}. This new proposal showed that its action expanded to the second order is free of ghost instabilities and discussed its possible embedding in a more fundamental theory.

Other works explore possible gravitomagnetic effects in galaxy rotation curves from linearized gravity and the post-Newtonian formalism \cite{Ludwig2021, Govaerts2023, Glampedakis2023, Lasenby2023}. For instance, Ludwig \cite{Ludwig2021} reproduces galactic rotation curves by using a non-linear differential equation that relates the rotation velocity to the mass density. Specifically, the author proposes a bidimensional model to fit total mass, central mass density and overall shape of the galaxies, according to observed luminosity and rotation curves. A recent study of Govaerts \cite{Govaerts2023} also obtains flat velocity curves for some especial configurations: Let $J$ be the angular momentum of a gravitational system with mass $M$, for positive Newman-Unti-Tamburino charges, values of $\frac{1}{4}|J|/M^2 \sim 9\cdot 10^5$ lead to plateau velocities of about $v/c \sim 8\cdot 10^{-4}$ for distances ranged in $400~000 < r/(\mathrm{G}M) < 1~500~000$. %%%% These velocities corresponds to $\mathrm{log}_{10}(v/[ms^{-1}]) \sim 4.4$, which is compatible with observations in flat rotation curves. %%% distances $r >> \mathrm{G}M > 0$ 

However, Glampedakis and Jones \cite{Glampedakis2023} state that post-Newtonian effects are too small to contribute significantly to the observed flat rotation curves. Similarly, Lasenby et al. \cite{Lasenby2023} show that gravitomagnetic effects on the circular velocity $v$ of a star are smaller than the standard Newtonian effects and thus any additional contribution to the galaxy rotation curves must be negligible.

\subsection{A natural relativistic formulation?}

The present work aims to formulate MOND-like dynamics with the minimum possible number of Lagrangian terms using natural features of hyperconical universes in agreement with our previous work \cite{Monjo2017, Monjo2018, MCS2020, MCS2023}. To address this goal, a hyperconical model was built by concatenating two steps. First, the use of moving frames produces a radial inhomogeneity from the extrinsic viewpoint of the expansion. Second, stereographic projections assimilate the radial inhomogeneity as an apparent acceleration under the intrinsic perspective.

Specifically, let homogeneous universes have positive ($k>0$), null ($k \approx 0$), or negative ($k<0$) curvature. Thus, moving frames in linearly-expanding homogeneous universes lead to a radial inhomogeneity in the isotropic spacetime \cite{Monjo2017}, whose differential line element is locally \begin{equation}
%ds^2 \approx dt^2 \left(1+\frac{2}{k}\sqrt{1- kr'^2} -\frac{2}{k} \right) 
ds^2 \approx dt^2 \left(1- kr'^2 \right) 
-  \frac{t^2}{t_{0}^2} \left( \frac{dr'^2}{1-kr'^2} + {r^{\prime}}^2d{\Sigma}^2 \right)
-  \frac{ 2r't}{t_{0}^2} \frac{dr'dt}{\sqrt{1-kr'^2}}
\end{equation}where $r' \ll t_0$ is the comoving distance, $\Sigma$ represents the angular coordinates and $t_0 \equiv 1$ is the current value for the age $t$ of the universe. Both the Ricci scalar of curvature and Friedmann equations derived from this universe for $k = 1 = 1/t_0^2$ are locally equivalent to those obtained for a flat $\Lambda$CDM model with linear expansion.
%with $c \equiv 1$

Observational compatibility of both extrinsic and intrinsic viewpoints was checked with Type Ia supernovae data \cite{Monjo2017}, ascertaining that the intrinsic measurement fits better (with $k>0$) when the luminosity distance is used to analyze redshifts. Theoretical compatibility between the expansion derived from the hyperconical model and the standard $\Lambda$CDM model was checked in \cite{Monjo2018}. In particular, compatibility demonstrates that there exists a projection that assimilates radial inhomogeneity as an apparent acceleration in expansion, such as the $\Lambda$CDM model. A particular solution (projection) was detailed in \cite{MCS2023}.

Arnowitt-Deser-Misner (ADM) formalism can be applied to analyze dynamics in modified gravity theories using their Lagrangian density and the corresponding Hamiltonian equations \cite{Arnowitt2008, Deruelle2010}. However, some authors do dispute the equivalence between ADM and GR due to differences in diffeomorphism invariance, since they are connected by non-canonical transformations \cite{Kiriushcheva2008, Salisbury2020}.

Assuming that GR and ADM are valid at least at a local scale, the consistency of the hypercone-based dynamics displayed three remarkable results in \cite{MCS2020}: (1) the total energy density of the universe is inactive since the equation of state is $w = -\frac{1}{3}$; (2) the spatial curvature of the hyperconical manifold needs to be $k = 1$; and (3) the Lagrangian density $\mathcal{L}$ requires a slight modification to \begin{equation}\label{lagrangian0}
\mathcal{L} = \frac{1}{16\pi\mathrm{G}} \Delta R +
\mathcal{L}_{M}
\end{equation}where $\mathcal{L}_{M}$ is the matter Lagrangian density and $\Delta R=R-R_u$ is the difference between the Ricci scalar curvature of the gravitational system and the Ricci scalar curvature of the background metric (hyperconical universe), $R_u$. Locally, if $k=1$, it is $R_u = -1/t_0^2$, with age $t_0$ of the universe. The modification of the Lagrangian density implies that metrics of the universe are globally independent of the matter content. Consistently, applying Friedmann equations to hyperconical cosmology \cite{MCS2020}, it is displayed that linear expansion produces the same equation of state of $w = -\frac{1}{3}$ that implies a globally “zero active gravitational mass-energy”.

Moreover, the solution of $k=1$ predicts \cite{MCS2023}: (1) a third-order $\Lambda$CDM-compatible (apparent) acceleration when the hyperconical universe is projected to an intrinsic viewpoint; (2) consequently, an apparent dark energy about $\Omega_\Lambda = 0.70$ is found as a constant for any age of the universe; (3) dark matter is another parameter whose value could be totally or partially consequence of the distortion projection; and (4) the Hubble tension observed between direct geometrical methods (e.g., distance ladder) and indirect methods (e.g., CMB) can be explained by the difference between the extrinsic and intrinsic viewpoints of the hyperconical universe.

% Spitzer Photometry and Accurate Rotation Curves (SPARC) database (Lelli et al. 2016) contains 175 rotationally supported galaxies in the nearby universe.

The present work assumes that GR needs to be adapted to our previous results \cite{Monjo2017, Monjo2018, MCS2020, MCS2023}, especially concerning the `zero active mass' ($w = -\frac{1}{3}$), also proposed by Melia \cite{Melia2017}, and the modified Lagrangian density (Eq. \ref{lagrangian0}). To reach the goal of deriving natural galaxy dynamics, this paper is structured in four main sections: Sec. 2 describes the theoretical framework of the hyperconical model; Sec. 3 details the derivation of rotation curves; Sec. 4 analyzes observational constraints with galaxies collected from the SPARC database \cite{Lelli2016b, Lelli2019}, and Sec. 5 enumerates the main conclusions of the work.

%DATA
%http://astroweb.cwru.edu/SPARC/
%http://astroweb.cwru.edu/SPARC/MassModels_Lelli2016c.mrt
%https://www2.mpia-hd.mpg.de/THINGS/Data.html
%https://iopscience.iop.org/article/10.1088/0004-6256/136/6/2563/meta
%https://iopscience.iop.org/article/10.1088/0004-6256/136/6/2648#aj266574s2
%http://simbad.u-strasbg.fr/simbad/sim-ref?querymethod=bib&simbo=on&submit=submit+bibcode&bibcode=2016PASJ...68....2S
%https://ned.ipac.caltech.edu/uri/NED::InRefcode/2016PASJ...68....2S

%REFERENCES
%https://link.springer.com/article/10.1140/epjc/s10052-022-10506-7
%https://academic.oup.com/mnras/article/439/2/2132/1020095?login=false
%https://ui.adsabs.harvard.edu/abs/2016PASJ...68....2S/abstract

%Rievers, B.; Lämmerzahl, C. (2011). "High precision thermal modeling of complex systems with application to the flyby and Pioneer anomaly". Annalen der Physik. 523 (6): 439. arXiv:1104.3985. Bibcode:2011AnP...523..439R. doi:10.1002/andp.201100081.

%thermal recoil force (TRF) is the accepted explanation for the Pioneer anomaly
%https://link.springer.com/article/10.12942/lrr-2012-10

\section{Theoretical framework}

\subsection{Hyperconical model}

The model used in this study is the hyperconical universe, which we developed in \cite{Monjo2017, Monjo2018, MCS2020}. The basic idea of the model is that the universe is globally homogeneous under an extrinsic viewpoint, but not under its intrinsic metric at large scales. Both perspectives are key points to develop the model: (1) the extrinsic curvature of the Universe and (2) the stereographic projection to intrinsically assimilate the linearly-expanding curved space as a flat with fictitious acceleration. Therefore, the model derivation needs to start from the minimal ambient space, which is a five-dimensional (flat) Minkowskian spacetime $(+,-,-,-,-)$, and obtaining the projective angles that allow an assimilation of the extrinsic linearly-expanding curvature as a radial inhomogeneity and then as a fictitious acceleration.

Our model of the universe is built on a Hypercone $\mathcal{H}^4$ embedded in a $\mathbb{R}^5$-Minkowskian spacetime with globally hyperbolic (Cauchy) hypersurfaces representing the space. That is a combination of a 3-sphere $S_1^3 \subset \mathbb{R}^4$ and a timeline $\mathbb{R}_{+} := \mathbb{R}_{>0} := (0, \infty)$, expressed as\begin{equation}\nonumber
\mathcal{H}^4 := S_{\mathbb{R}_{+}}^3 := \{q = (t, \vec \rho) \in \mathbb{R}^5, \; | \; \vec \rho \in S_{t}^3, \, t \in \mathbb{R}_{+}\} \subset \mathbb{R}^5,
\end{equation}with metric $\eta_{1,4} = \mathrm{diag}(1,-1,-1,-1,-1)$. That is, coordinates of a point $q \in \mathcal{H}^4$ are chosen here as $q  = (t, \vec \rho):= (t, \vec{r}, u) := (t, x, y, z, u)$ with $\vec \rho^2 = \vec{r}^2 + u^2 = x^2 + y^2 + z^2 + u^2 = t^2$, or simply $q^2 = 0$.

\textbf{Line element}. Since observers should measure the same \textit{length element} in a flat (Minkowskian) manifold as in a local tangent space of our universe, their expansion needs to be locally removed from the extrinsic frame to obtain an observed-intrinsic viewpoint of its metric (similar to Eq. \ref{lagrangian0}). Consequently, according to Monjo \cite{Monjo2018} as well as Monjo and Campoamor-Stursberg \cite{MCS2023}, the extrinsic homogeneous positive curvature (global symmetry) of the spatial part $\mathcal{H}^4|_{t=t_0}$ is broken when an observer measures ($t<t_0$)-regions at the time $t_0 \in \mathbb{R}_{+}$. Since any inertial observer is comoving to the expansion, a non-isometric transformation is required to find the intrinsic perspective of the observer (actually, of the light observed). Specifically, let $s : \mathbb{R}_{+} \to \mathcal{H}^4$ be the path of an observer and $s(t) := (t,\vec 0, t)$ be its coordinates at the time $t$; therefore, the line element of the background metric of the hyperconical universe is \begin{equation} \label{eq:exact_line}
ds_{\mathcal{H}^4}^2 \;\;  = \;\; dt^2 \left(2 \sqrt{1-\frac{{ r^{\prime}}^2}{ t_{0}^2}} - 1 \right) 
-  \frac{t^2}{t_{0}^2} \left( \frac{d{r^{\prime}}^2}{1-\frac{{r^{\prime}}^2}{t_{0}^2}} + {r^{\prime}}^2d{\Sigma}^2 \right)
-  \frac{{ 2r^{\prime}t}}{t_{0}^2} \frac{dr^{\prime}dt}{\sqrt{1-\frac{{r^{\prime}}^2}{ t_{0}^2}}}\;\;
%\\  \label{eq:exact_line}
%& = & dt^2 \left(2 \sqrt{1-\frac{{ r^{\prime}}^2}{ t_{0}^2}} - 1 -  \frac{ 2r't}{t_0^2} \frac{\dot r'}{\sqrt{1-\frac{{r^{\prime}}^2}{ t_{0}^2}}} \right) 
%-  \frac{t^2}{t_{0}^2} \left( \frac{d{r'}^2}{1-\frac{{r^{\prime}}^2}{t_{0}^2}} + {r^{\prime}}^2d{\Sigma}^2 \right) \,\;\;\hspace{5mm}
\end{equation} Here, spherical coordinates $(dt, dr, r\,d{\Sigma})$ are replaced by the comoving ones ($dt$, $a(t)\,dr'$, $a(t)r'd{\Sigma}$) with a linear expansion factor $a(t) := t/t_{0}$ such that $r'/t_0 = t/t$ and a solid angle $d\Sigma := \sin \theta \,d\theta \,d\varphi$, which can be built with angular coordinates such as $(\theta, \varphi) \in [0,\pi] \times [0, 2\pi)$. For the first-order approach, the line element yields to: \begin{equation} \label{eq:approx1}
ds_{\mathcal{H}^4}^2 \approx dt^2 \left(1-\frac{ r'^2}{ t_{0}^2}  \right) 
-  \frac{t^2}{t_{0}^2} \left( \frac{d{r'}^2}{1-\frac{{r'}^2}{t_{0}^2}} + {r'}^2d{\Sigma}^2 \right) -  2\frac{t}{t_0} \frac{ r'}{t_0} dr'dt  
%%%% -  2\dot r'\frac{t}{t_0} \frac{ r'}{t_0}
\end{equation} 

\textbf{Vacuum cosmology}. In the case of an (unperturbed) homogeneous universe, linear expansion of ${\mathcal{H}^4}$ can be expressed in terms of the vacuum energy density,  $\rho_0 = \rho_{crit} =
3/(8\pi\mathrm{G}t^2)$, where $\mathrm{G}$ is the Newtonian gravitational constant. That is, one can define an inactive (vacuum) mass or energy $\mathcal{M}(r) =\rho_0 \frac{4}{3}\pi r^3$ for a distance equal to $r$ with respect to the \textit{reference frame origin}. By using the relation between the original coordinates $(dt, dr, r\,d{\Sigma})$ and the comoving ones $(dt, a(t)\,dr', a(t)r'd{\Sigma})$, the spatial dependence of Eq. \ref{eq:approx1} is now \begin{equation} \label{eq:metric_GM}
\frac{{r'}^2}{t_{0}^2} = \frac{{r}^2}{t^2} = \frac{2\mathrm{G}\rho_0\frac{4}{3}\pi r^3}{r}  = \frac{2\mathrm{G}\mathcal{M}(r)}{r} \,,
\hspace{10mm} 2 \frac{t}{t_0} \frac{ r'}{t_0}  \; = \;  2 \frac{t}{t_0} \sqrt{\frac{2\mathrm{G}\mathcal{M}(r)}{r}}\,.
\end{equation} Then, Eq. \ref{eq:approx1} yields to \begin{eqnarray} \nonumber
ds_{\mathcal{H}^4}^2 & \approx & \left(1- \frac{2\mathrm{G}\mathcal{M}(r)}{r} \right) dt^2
-  \frac{t^2}{t_{0}^2} \left[\left(1- \frac{2\mathrm{G}\mathcal{M}(r)}{r}\right)^{-1}d{r'}^2 + {r'}^2d{\Sigma}^2 \right] + \\
&& + 2 \frac{t}{t_0} \sqrt{\frac{2\mathrm{G}\mathcal{M}(r)}{r}}
\end{eqnarray}Considering $\dot r' \ll r'/t_0$ for the shift term and also for  $dr'^2 = \dot r'^2 dt^2 \ll  r'^2/t_0^2 dt^2$, \begin{equation} \label{eq:approx2}
ds_{\mathcal{H}^4}^2 \bigg|_{\dot r' \ll r'/t_0} \approx  \left(1- \frac{2\mathrm{G}\mathcal{M}(r)}{r} \right) dt^2
-  \frac{t^2}{t_{0}^2} \, {r'}^2d{\Sigma}^2\,.
\end{equation} In any case, only the component $g_{tt}$ significantly contributes to the local limit of gravity fields.

\begin{definition}[Mass of perturbation] From the modified Lagrangian density displayed in Eq. \ref{lagrangian0}, a perturbation of the vacuum density $\rho_0 \to \rho_M(r) := \rho_0 + \Delta \rho$ leads to a mass $M := \frac{4}{3}\pi r^3 \Delta \rho$, which is likewise obtained by perturbing the curvature term, $r^2/t^2 \to r^2/t_M(r)^2 := r^2/t^2 + 2\mathrm{G}M/r$ with the radius of curvature $t_M \le t$ (see  more details on its related Lagrangian density in \textbf{Appendix A}).\end{definition}

Then, according to the above definition, Eq. \ref{eq:approx2} becomes a line element with a temporal component similar to the Schwarzschild case (see \textbf{Appendix B})\begin{equation} \label{eq:Schwarzschild}
ds_{\mathcal{H}^4}^2 \bigg|_{\dot r' \ll r'/t_0}  \approx  \left(1- \frac{r^2}{t^2} - \frac{2\mathrm{G}M}{r} \right) dt^2
-  \frac{t^2}{t_{0}^2}\, {r'}^2d{\Sigma}^2 \;,
\end{equation} whose metric $g_{\alpha\beta} = \eta_{\alpha\beta} + h_{\alpha\beta}$ has a slightly different perturbation $ h_{\alpha\beta}$, such as $h_{tt} \approx - \frac{r^2}{t^2} - \frac{2\mathrm{G}M}{r}$. In any way, Eq. \ref{lagrangian0} states that the background curvature of the universe does not produce gravitational effects and thus it can be neglected. %%and $h_{r'r'} \approx h_{tt}$

\subsection{Projective angles}

For closed and homogeneous universes, without density perturbations nor projective transformations, all inertial velocities $v$ are due to the expansion of the universe like in an empty spacetime. For each angle $\gamma$, the Hubble law is simply $v = H r = \frac{r}{t} = c \sin \gamma$, with Hubble parameter $H = \frac{1}{t}$ and a spatial domain $\gamma \in [0,\gamma_U)$ of the exact line element (\ref{eq:exact_line}), satisfying $\sin\gamma_U = \sqrt{3/4}$, which is $\gamma_U = \frac{1}{3}\pi \sim 1$ \cite{Monjo2017,Monjo2018}. This is the projective domain for an empty hyperconical universe with positive curvature $k = 1$, specifically $\gamma_U := \sin^{-1} \sqrt{1-k/4} = \pi/3$ \cite{Monjo2018}. Then, projecting the (positively-curved) hyperconical universe onto a plane, the resulting projected angle is $\gamma \in [0, \gamma_0)$ with a maximum $\gamma_0$ that is double $\gamma_U$  (\cite{MCS2023}):\begin{equation} \label{eq:projective0}
\gamma_0(\gamma_U)  \approx \frac{\gamma_U}{\cos \gamma_U} \sim \frac{2}{3}\pi \sim 2\,.
\end{equation}See derivation details in \textbf{Appendix C}. After the projection, our model becomes theoretically compatible with the standard $\Lambda$CDM model up to the third order of approaching (providing the same cosmological parameters of \{$\Omega_\Lambda \approx 0.7$, $\Omega_m \approx 0.3$, $\Omega_K \approx 0$\}) for every time $t_0$ \cite{MCS2023}. That is, the dark energy and matter are interpreted as apparent quantities for the observers since the observed paths of light are intrinsic in the manifold. Nevertheless, the cosmic timeline is the same as in the extrinsic perspective, with a linear expansion and `zero active mass' ($w = -\frac{1}{3}$) \cite{MCS2020}.

The projective angle defined in Eq.  \ref{eq:projective0} changes when vacuum energy is perturbed. In such a case, the gravitational system (Eq. \ref{eq:Schwarzschild}) results in a characteristic scale of $r_{cs}(M) := t_M(r) \sin \gamma_M(r)$ given by a maximum angle $\gamma_M \ge \gamma_U := \frac{\pi}{3}$ that is approximately constant, $\gamma_M \in [\pi/3, \pi/2)$, but it slightly depends on the radial distance $r$ and on a function of the mass $M$. Since the gravitational system perturbs the cosmological geometry with an escape speed of $v_E^2(r) = \frac{2\mathrm{G}M}{r}$, the following is expected: \begin{equation}
    \sin^2 \gamma_M(r) = \frac{r_{cs}^2(M)}{t_M^2(r)} = \frac{r_{cs}^2(M)}{t^2} + \frac{2r_{cs}^2(M)\, \mathrm{G}M}{r^3} = \sin^2\gamma_U + \beta(r)\frac{2\mathrm{G}M}{r}\,,
\end{equation} with $\beta(r) := r_{cs}^2/r^2 \gg 1$, which is $\sin^2 \gamma_M (r) \approx \sin^2  \gamma_U + \beta(r) v_E^2(r)$ by approaching $\sin^2  \gamma_U \approx r_{cs}^2/t^2$ when $2\mathrm{G}M \ll r_{cs}(M) \sim t$.  On the opposite side, $\sin^2 \gamma_M \sim \sin^2 \gamma_{gc} \sim 1$ for regions close to the galaxy center (angle $\gamma_{gc}$ when $r = 2\mathrm{G}M  \ll t$), but with small cosmological perturbations on the surroundings as\begin{equation} \label{eq:gamma_gc}
   1 \approx \sin^2 \gamma_{gc} \approx \frac{r_{cs}^2(M)}{t_M^2(r)} + \frac{r_{cs}^2(M)}{3\,t^2} \approx \frac{4}{3}\beta(r)\frac{r^2}{t^2} + \beta(r)\frac{2\mathrm{G}M}{r}\,.
\end{equation} Since $t_M^2(r) \in (4\mathrm{G}^2M^2, t^2]$ and  $r_{cs}^2(M) \in (4\mathrm{G}^2M^2, \frac{3}{4}t^2]$, $\,r_{cs}^2(M)$ is increasing from $r_{cs}^2(M)=t_M^2(r)=4\mathrm{G}^2M^2 \ll t^2$ up to $r_{cs}^2(M)=\frac{3}{4}t_M^2(r)=\frac{3}{4}t^2 = t^2\sin^2\gamma_U$, reducing the maximum angle from $\gamma_M = \frac{\pi}{2}$ to $\gamma_M = \frac{\pi}{3}$. At galactic scales, Eq. \ref{eq:gamma_gc} can be approached by $\sin^2 \gamma_{gc} \sim \sin^2 \gamma_M(r) + \beta(r) v_H^2(r) \sim  \sin^2  \gamma_U + \beta(r) v_E^2(r)  + \beta(r) v_H^2(r)$ and, applying the quotient to remove the dependency on $\beta(r)$, it is \begin{equation} \label{eq:model_g0}
 \frac{\sin^2 \gamma_M(r) - \sin^2 \gamma_U}{\sin^2\gamma_{gc} -  \sin^2 \gamma_U} \sim \frac{v_E^2(r)}{v_H^2(r) + v_E^2(r)}\,.
\end{equation}  Two limiting cases are $\sin \gamma_M \approx 1$ when $v_E(r) >> v_H(r)$ and  $\sin \gamma_M \approx \frac{\sqrt{3}}{2}$  when $v_E(r) \ll v_H(r)$.  Summarising, $\gamma_U$ and $\gamma_{gc}$ are assumed to be constant, while $\gamma_M$ (or $\gamma_0$) depends on the energy density perturbation.

\subsection{Cosmological projection}

Henceforth, the constant of the light speed $c\equiv 1$ will be not omitted in the equations to compare with real observations later. Let $\lambda_u$ be the scaling factor of a stereographic projection (\textbf{Appendix C}) of the coordinates $(r, u) = (ct\sin\gamma,\; ct\cos\gamma)$, \begin{equation} \label{eq:lambda_global0}
\lambda_u  \approx  \frac{1}{1-\frac{\gamma}{\gamma_0}}\,,
\end{equation} where $\gamma_0 = \gamma_U/\cos \gamma_U = \frac{2}{3}\pi$, since empty spacetimes have $\gamma_M = \gamma_U$. For non-empty matter densities, we contend that $\gamma_M$ depends on the escape speed of the gravity system considered, assuming that dependency on distances is weak, thus $\gamma_0$ is approximately constant for each case.  The spatial distortion is given by an exponent $\alpha > 0$ in the scaling factor $\lambda$, such as: \begin{equation} \label{eq:lambda_globa2}
\lambda_u^{\alpha}  \approx   {1+\frac{\alpha \gamma}{\gamma_0}} \approx {1+\frac{\alpha r'}{\gamma_0 t_0c}} \,.
\end{equation} Therefore, the coordinates are \begin{eqnarray} \label{eq:lambda_globa3}
\hspace{24mm} 
\hat r'  &=& \lambda_u^{\alpha} r' \approx \left( {1+\frac{\alpha r'}{\gamma_0 t_0c}} \right) r' \\
\hspace{24mm} 
\hat t &=& \lambda_u t \approx \left(1 +  \frac{r'}{\gamma_0t_0c}\right) t
\end{eqnarray} At a local scale, $\alpha = 1/2$ is required to guarantee consistency in dynamical systems \cite{MCS2023}, but the parameter $\alpha$ is not essential in this work.

Applying this projection to the metric and the corresponding geodesics, it is easy to obtain a first-order approach of the cosmological perturbation that contributes to modify the Newtonian dynamics in the classical limit.

%%%
 \subsection{First-order perturbed metric}

%https://physics.stackexchange.com/questions/308303/vierbeins-in-schwarzschild
Assuming that $\gamma_0$ is approximately constant, the differential and quadratic form of the time coordinate displays
\begin{eqnarray} \label{eq:lambda_globa4} \nonumber
%%% d \hat r' &\approx &  \left(1 +  \frac{\alpha r'}{\gamma_0 t_0c}\right)  dr'  
%%% \\  \nonumber
d \hat t &\approx & \left(1 +  \frac{r'}{\gamma_0 t_0c}\right)dt +  \frac{t}{\gamma_0 t_0c} dr'  
%%% \\ \nonumber
%%% d \hat r'^2 &\approx& \left(1 +  \frac{2\alpha r'}{\gamma_0 t_0c}\right)  dr'^2 
\\ \nonumber
  {d\hat t}^2  &\approx & \left(1 +  \frac{r'}{\gamma_0 t_0c}\right)^2 dt^2 + \left(\frac{t}{\gamma_0 t_0c}  \right)^2 dr'^2 
+2\left(1 +  \frac{r'}{\gamma_0 t_0c}\right)\frac{t}{\gamma_0 t_0c}\frac{dr'}{dt} dt^2 \approx 
\\
& \approx & \left(1 +  \frac{2 r'}{\gamma_0 t_0c} + \frac{2t\dot r'}{\gamma_0 t_0c}  \right) dt^2 \;\;+\;\;\textrm{upper-order terms}\,.
\end{eqnarray} By using these prescriptions, the modified Schwarzschild metric (Eq. \ref{eq:Schwarzschild}) is projected $d s_{\mathcal{H}^4}^2  \to  d \hat s^2 $, then the new coordinates $(\hat t, \hat r')$ are expressed in terms of the original ones $(t, r')$, and finally, it is locally expanded up to first-order perturbations terms: \begin{eqnarray} \nonumber
d \hat s^2 &\approx&  \left(1- \frac{ \hat r^2}{ c^2\hat t^2} - \frac{2\mathrm{G}M}{c^2 \hat r} \right) c^2 d \hat t^2
-  \frac{ \hat t^2}{t_{0}^2} \, { \hat r'^2}d{\Sigma}^2
%\frac{ \hat t^2}{t_{0}^2} \left[\left(1- \frac{  \hat r^2}{c^2 \hat t^2} - \frac{2\mathrm{G}M}{c^2 \hat r}\right)^{-1}d{ \hat r'^2} + { \hat r'^2}d{\Sigma}^2 \right] \;\; \;\; %  \\ \nonumber & \approx & 
  \;\; \approx \;\;   \hat g_{tt}c^2dt^2 + \hat g_{ii}(dx^i)^2 \,.
\end{eqnarray} Notice that, according to  Eq. \ref{lagrangian0}, the background terms $r'^2/t_0^2$ do not produce gravitational effects and thus they can be neglected. Here, one identifies a projected perturbation $\hat h_{tt}$ of the temporal component of the metric, $\hat g_{tt} = \eta_{tt} + \hat h_{tt} = 1 + \hat h_{tt}$. Thus, if $M$ is the mass of a central gravity source, the first-order perturbation of the temporal component of the metric is \begin{eqnarray} \nonumber \label{eq:gtt}
\hat h_{tt} &\approx& \underbrace{- \frac{r^2}{c^2t^2}\left( {1+\frac{\alpha r'}{\gamma_0 t_0c}} \right)^2}_{\mathrm{neglected}} - \frac{2\mathrm{G} M}{rc^2}\left({1-\frac{\alpha r'}{\gamma_0 t_0c}} \right) +  \frac{2r'}{\gamma_0 c t_0} + \frac{2t{\dot r'}}{\gamma_0 t_0c} \approx \\ \nonumber
&\approx&  - \frac{2\mathrm{G} M}{rc^2}\left({1-\frac{\alpha r}{\gamma_0 t c}} \right)  +  \frac{2}{\gamma_0 c} \left(\frac{r}{t} + \frac{t}{ t_0} {\dot r'} \right) %% \approx - \frac{2\mathrm{G} M}{rc^2}  +  \frac{2r}{\gamma_0 tc}
%=: 1 - \frac{2\mathrm{G}}{c^2  r}\left(M - \frac{2}{\gamma_0}\frac{E_{rt}}{c^2}\right)
\\ \nonumber 
\frac{\partial}{\partial t} \hat h_{tt} &\approx&  - \frac{2\mathrm{G} M}{rc^2}\left({\frac{\alpha r}{\gamma_0 t^2 c}} \right)  +  \frac{2}{\gamma_0 c} \left(-\frac{r}{t^2} + \frac{1}{ t_0} {\dot r'} \right) 
\approx
\frac{2\alpha \mathrm{G} M}{\gamma_0 t^2c^3}   +  \frac{2}{\gamma_0 c} \left(\frac{\dot r'}{ t_0}  -\frac{r}{t^2} \right)  
\\ \nonumber 
\frac{\partial}{\partial r} \hat h_{tt} &\approx&
 - \frac{2\mathrm{G} M}{r^2 c^2}  +  \frac{2}{\gamma_0 t c} 
%\approx -   \frac{2\mathrm{G} M}{r^2 c^2}  +  \frac{2}{\gamma_0 t c} - \frac{2r}{\gamma_0^2 t c}\frac{d\gamma_0}{dr}  
%\approx - \frac{2\mathrm{G} M}{r^2 c^2} +  \frac{2}{\gamma_0 t c}    + \mathcal{O}(\gamma_0')   
\end{eqnarray} where the relation between the comoving distance $r'$ and the spatial coordinate $r$ is used, that is $r'/t_0 = r/t$.

\subsection{First-order perturbed geodesics}

%https://www.vttoth.com/CMS/physics-notes/312-the-newtonian-limit-in-general-relativity
%http://einsteinrelativelyeasy.com/index.php/general-relativity/38-newtonian-limit

Under the Newtonian limit of the GR, the largest contribution to the gravity dynamics is given by the temporal component of the metric perturbation $\hat h_{tt}$. That is, the Schwarzschild geodesics, 
\begin{equation} \label{eq:geodesic0}
\frac{d^2x^\mu}{d\tau^2} \approx \frac{1}{2}\eta^{\mu\nu} \frac{\partial}{\partial x^\nu} \hat h_{tt} \left(\frac{cdt}{d\tau}\right)^2 \;,
\end{equation} is perturbed by the distorted steoreographic projection.

On the right side of Eq. \ref{eq:geodesic0}, one finds time-like and space-like components from the metric perturbation $\hat h_{tt}$, with a (flat) metric signature $\eta_{\mu\nu} = \eta^{\mu\nu}= \mathrm{diag}(1,-1,-1,-1)$, \begin{equation}
\frac{d^2 \hat s}{d\tau^2} \approx  \left(\frac{1}{2} \frac{\partial}{\partial x^0} \hat h_{tt} \,e_t \; - \;  \frac{1}{2} \frac{\partial}{\partial x^i} \hat h_{tt}\,e_i  \right) \left(\frac{cdt}{d\tau}\right)^2\,,
\end{equation} where the four-position $\hat s := (c\Delta t, x^i) = c \Delta t\, e_t \,+\, x^i\, e_i =: c \Delta\mathbf{t} + \mathbf{x} \in \mathbf{R}^{1,3}$ is assumed.

Considering a Schwarzschild metric with cosmological perturbation for a test particle orbiting around a central mass $M$, its geodesic equation is \begin{eqnarray}  \nonumber
\frac{d^2 \hat s}{c^2dt^2} &\approx & \frac{1}{2} \frac{\partial \hat h_{tt}}{c\partial t}\,e_t \; - \;  \frac{1}{2} \frac{\partial \hat h_{tt}}{\partial x^i} \,e_i    =  \frac{1}{2} \frac{\partial \hat h_{tt}}{c\partial t}   \,e_t \; - \; \frac{1}{2} \frac{\partial \hat h_{tt}}{\partial r} \frac{\partial r}{\partial x^i} \,e_i  
\\ \label{eq:geodesic1}
\frac{d^2 \hat s}{c^2dt^2} &\approx& -  \left( \xcancel{- \frac{{\dot r'}}{\gamma_0 t_0c^2}} + \frac{r}{\gamma_0 c^2t^2} + \frac{\alpha \mathrm{G} M}{\gamma_0 t^2c^4}   \right)  \,e_t \; - \;  \left(\frac{\mathrm{G}M}{c^2r^2}+\frac{1}{\gamma_0 ct}\right)\frac{x^i}{r} \,e_i
\end{eqnarray} where the first term of the right side is \textbf{neglected for orbits}. This is a simple space-like vector, but for $r/t \ll 1$, it is mostly determined by the Newtonian acceleration $a_N := -\frac{\mathrm{G}M}{r^2}\,\frac{x^i}{r} \,e_i$ within the spatial component. For a free-fall particle with central-mass reference coordinates $\mathbf{x} = x^ie_i = (r, 0, 0) = \mathbf{r} \in \mathbb{R}^3$, it experiences a spatial acceleration equal to\begin{equation}\label{eq:Pioneer0}
\mathbf{a} := a^re_r := \frac{d^2 \mathbf{r}}{dt^2} \approx \; - \;  \left(\frac{\mathrm{G}M}{r^2}+\frac{c}{\gamma_0 t}\right)\frac{\mathbf{r}}{r} \; = \;  a_N  - \frac{c}{\gamma_0 t}\frac{\mathbf{r}}{r}\,.
\end{equation}That is, an acceleration anomaly is obtained in the spatial direction, about $|\mathbf{a} - a_N| \approx \gamma_0^{-1}c/t$. However, total acceleration also has a time-like component, that is, in the direction $e_t$.

\section{Deriving rotation curves}

 \subsection{Hypercone-based discrepancy model}

\subsubsection{Time-like contribution to the centrifugal acceleration.} By using the Eq. \ref{eq:geodesic1} plus additional assumptions, it is possible to derive an expression for rotation velocity curves of galaxies and their mass-discrepancy acceleration relationship. The assumptions we are referring to are: i) extrinsic curvature of our universe contributes to centrifugal acceleration in the time-like direction, and ii) for orbital velocity, the variation of radial distance can be approximated to zero, which is $\dot r' = dr'/dt \approx 0$, compared to the other contributions. On the left side of Eq. \ref{eq:geodesic1}, centrifugal acceleration has also both time-like and space-like components in the mostly spatial radius direction, $e_s := \hat s /||\hat s|| = - e_t \sinh\gamma - e_{\vec r}  \cosh\gamma$, where $\hat s$ is the position vector with respect to the central mass $M$ and $e^te_t = -e^{\vec r}e_{\vec r} = -e^s e_s = 1$ since the metric signature is $(+, -, -, -)$. Notice that the (new) coordinates in the direction $e_s$ are given by a Lorentz rotation with a relative speed of the Hubble law $v_H := \frac{r}{t} = c\sin\gamma$, that is, $\gamma = \arcsin(\frac{r}{ct}) = \arcsin(\frac{r'}{ct_0})$.

Thus, time-like contribution to the centrifugal acceleration is  nonzero and it is  proportional to $1/ct$ as the spatial contribution is proportional to $1/r$. Using these ingredients, centrifugal acceleration is rewritten in terms of the spatial velocity $v$ and the acceleration direction $e_s$, as follows:
\begin{eqnarray} \nonumber
\hat{a} &:= &\frac{d^2{\hat s}}{c^2dt^2} = -\omega_t^2 ct \, e_t \sinh\theta - \omega_r^2 r\, e_{\vec r}  \cosh\theta  = \\ \label{eq:centrif1}
&=& -\left(\frac{1}{ct} e_t e^t + \frac{1}{r} e_{\vec r} e^{\vec r}\right)\frac{v^2}{c^2} {e_s} =: S^{-1} \frac{v^2}{c^2} {e_s} \,,
\end{eqnarray} where $\omega_t := v/(ct)$ and $\omega_r := v/r$ are the angular speeds in the time- and space-like directions, respectively, assuming a tangential speed $v$. 

\subsubsection{Mass-discrepancy acceleration curve.} By using the assumption of time-like direction contribution to the centrifugal acceleration, a mass-discrepancy acceleration relation can be derived. Applying the inverse $S$ of the diagonal matrix $S^{-1} := -\left(\frac{1}{ct} e_t e^t + \frac{1}{r} e_{\vec r} e^{\vec r}\right)$ to the Eq. \ref{eq:centrif1}, in this way \begin{eqnarray} \nonumber
\frac{v^2}{c^2} {e_s} &=& -\left(ct e_t e^t + x^i e_i e^i\right) \frac{d^2{s}}{c^2dt^2} 
\approx \\
& \approx &
 \,ct\left( \frac{\alpha \mathrm{G} M}{\gamma_0 t^2c^4}  + \frac{r}{\gamma_0 c^2t^2}\right) \,e_t \; + \;  \left(\frac{\mathrm{G}M}{c^2r^2}+\frac{1}{\gamma_0 ct}\right)\frac{x^i x_i}{r} \,e_i\;,
\end{eqnarray}an effective space-like direction ($||e_s||^2 = e_se^s = -1$) is found, and the absolute value of the velocity is given by\begin{eqnarray} \nonumber
 \frac{v^4}{c^4} =
  - \bigg|\bigg| \frac{v^2}{c^2}  {e_s} \bigg|\bigg|^2 & \approx &  -  \left[ \left( \frac{\alpha \mathrm{G} M}{\gamma_0 tc^3} + \frac{r}{\gamma_0  c t}\right)^2 - \left(\frac{\mathrm{G}M}{c^2r} + \frac{r}{\gamma_0 c t} \right)^2  \right] =
\\ \nonumber
&\approx &\left(\frac{\mathrm{G}M}{c^2 r}\right)^2 +\frac{2\mathrm{G}M}{\gamma_0 t c^3} - \left(\frac{\alpha \mathrm{G} M}{\gamma_0 tc^3}\right)^2 - \frac{2\alpha \mathrm{G} M\, r}{\gamma_0^2 t^2c^4}
\\  \nonumber
&\approx & 
\left(\frac{\mathrm{G}M}{rc^2}\right)^2\left(1 - \frac{\alpha^2 r^2}{\gamma_0^2 c^2 t^2} \right) +\frac{2\mathrm{G}M}{\gamma_0 t c^3}\left(1 - \frac{\alpha r}{\gamma_0ct}\right) \\ 
& \approx & \left(\frac{\mathrm{G}M}{rc^2}\right)^2 +\frac{2\mathrm{G}M}{\gamma_0 t c^3}
\;, 
\end{eqnarray}which satisfies two well-known limits of Newton's dynamics and the Milgrom's MOND:\begin{eqnarray}
\hspace{20mm} 
v \approx \sqrt{\frac{\mathrm{G}M}{ r}}\;\;  & \hspace{10mm} & \mathrm{if}\; \frac{\mathrm{G}M}{r^2} >> \frac{2c}{\gamma_0 t} =:  a_0 \\
\hspace{20mm} 
v \approx \sqrt[4]{\frac{2\mathrm{G}Mc}{\gamma_0 t}}\;\;  & \hspace{10mm} & \mathrm{if}\; \frac{\mathrm{G}M}{r^2} \ll \frac{2c}{\gamma_0 t} =  a_0 \,,
\end{eqnarray} where $a_0$ is the Milogrom's acceleration parameter and $M = M(r)$ is the total mass within the central sphere of radius $r$. Finally, the velocity curve $v=v(r)$ can be reworded in terms of the  escape speed $v_E := \sqrt{2\mathrm{G}M_b(r)/r}$ or of the Kepler speed $v_K := \sqrt{\mathrm{G}M(r)/r}$ as follows:\begin{equation} \label{eq:MOND}
v^4  = v_K^4 + v_K^2 \, r \frac{2c}{\gamma_0 t} = \frac{1}{4}v_E^4 + v_E^2 \, r \frac{c}{\gamma_0 t} \;\;.
\end{equation} Therefore, the predicted mass-discrepancy acceleration relation is \begin{equation} \label{eq:discrepancy}
\frac{v^4}{ v_K^4}  = 1 + \frac{r}{v_K^2} \frac{2c}{\gamma_0 t} \;\; \Longrightarrow\;\; \left(\frac{v}{ v_K}\right)^2 = \sqrt{1 + \frac{1}{|a_N|} \frac{2c}{\gamma_0 t}}\;\;.
\end{equation} 

\subsubsection{MDAR/BTFR modeling.} 

The rotation curve is immediately obtained from Eq. \ref{eq:discrepancy}, which approaches to Eq. \ref{eq:Pioneer0} for $a_N >> c/t$ with $a_N = - v_K^2\,\mathbf{r}/ r^2$ and $d^2\mathbf{r}/dt^2 = - v^2 \mathbf{r}/ r^2$.  Another way to express the mass-discrepancy acceleration is the BTFR curve, which is an empirical function $v(M_b)$ between the baryonic mass ($M_b$) and the rotation velocity ($v$) for a certain radius $r$, which is an approximate constant value (i.e. a `flat curve'). In our case, that relationship is immediately found by:\begin{equation} \label{eq:BTFR}
 {v(r; \gamma_0)}  \approx  \sqrt[4]{ \left(\frac{\mathrm{G}M_b(r)}{r}\right)^2 +\frac{2\mathrm{G}M_b(r) c}{\gamma_0 t} } \;\;\Rightarrow\;\; {v(M_b; \gamma_0)}  \approx  \sqrt[4]{ \left(\frac{\mathrm{G}M_b}{r_{M_b}}\right)^2 +\frac{2\mathrm{G}M_b c}{\gamma_0 t} } \,,
 \end{equation}where $r_{M_b}$ is the galaxy radius as a function of the baryonic mass (e.g. following a similar idea to that used in stars \cite{Wu2018}). Notice that the first term of the root argument is the Newtonian contribution to the rotation curve, while the second term leads to the ``flatness'' of the rotation curve. To analyze the effect of the Newtonian contribution to the BTFR, we consider a constant angle $\gamma_0$ and an averaged radius (for all observed objects) in each galaxy.

\subsubsection{ $\gamma_{M}$-hypercone-based galaxy rotation curves. }

The characteristic angle $\gamma_M =\gamma_U$ for the empty regions is projected at $\gamma_0(\gamma_U) = \gamma_U/\cos \gamma_U = \frac{2}{3}\pi$.  However, nonzero perturbations locally modify the curvature of the spacetime and its causality angles, with additional movements to the expansion of the universe, and therefore $\gamma_0 \neq \frac{2}{3}\pi$ in most cases. In other words, it is expected that values of $\gamma_0$ depends on the difference between $v_E$ and $v_H$ according some relationship like Eq. \ref{eq:model_g0}. Thus, Eq. \ref{eq:projective0} becomes\begin{equation}
\gamma_0(\gamma_M) :\approx \frac{\gamma_M}{\cos \gamma_M} %% \;\;\Longrightarrow\;\;\gamma_0(r) :\approx \frac{\gamma_M(r)}{\cos \gamma_M(r)} 
\end{equation}  and, therefore, the orbital speed is now expressed in terms of the unique mean value of the parameter $\gamma_M$ as follows \begin{equation} \label{eq:BTFR2}
 {v(M_b; \gamma_M)}  \approx  \sqrt[4]{ \left(\frac{\mathrm{G}M_b}{r_M}\right)^2 + \frac{2\mathrm{G}M_b c}{t}\, \frac{\cos\gamma_M}{\gamma_M} } \end{equation} where the parameter $\gamma_M$ is assumed to be a constant.

\subsubsection{ $\gamma_{gc}$-hypercone-based galaxy rotation curves.}

Taking into account all possible values from the domain of the projective angles, the galactic center is ranged by $\gamma_{gc} \in (\frac{\pi}{3}, \frac{\pi}{2})$; thus, it is possible to model the characteristic angle $\gamma_M$ of Eq. \ref{eq:model_g0} as follows:  \begin{equation} \label{eq:model_g1}
\sin^2 \gamma_M(\gamma_{gc}) = \sin^2 \gamma_U + \left(\sin^2 \gamma_{gc}-\sin^2 \gamma_U\right)\frac{\frac{2\mathrm{G}M}{r}}{\frac{r^2}{t^2} + \frac{2\mathrm{G}M}{r}}\,. 
\end{equation} 
Thus, replacing $\gamma_M$ by its function, depending on $\gamma_{gc}$ in Eq. \ref{eq:BTFR}, a BTF relationship is obtained in terms of that parameter $v(M_b; \gamma_M) \Rightarrow  v(M_b, r; \gamma_{gc})$.

\begin{equation} \label{eq:BTFR3}
 {v(M_b, r; \gamma_{gc})}  \approx  \sqrt[4]{ \left(\frac{\mathrm{G}M_b}{r}\right)^2 +  \frac{2\mathrm{G}M_b c}{t}\, \frac{\cos\gamma_M(r,\gamma_{gc})}{\gamma_M(r,\gamma_{gc})} } \end{equation}\,.

\section{Observational constraints}

\subsection{Galactic scales}

\begin{figure}[pb]
\begin{center}
    \includegraphics[scale=0.75]{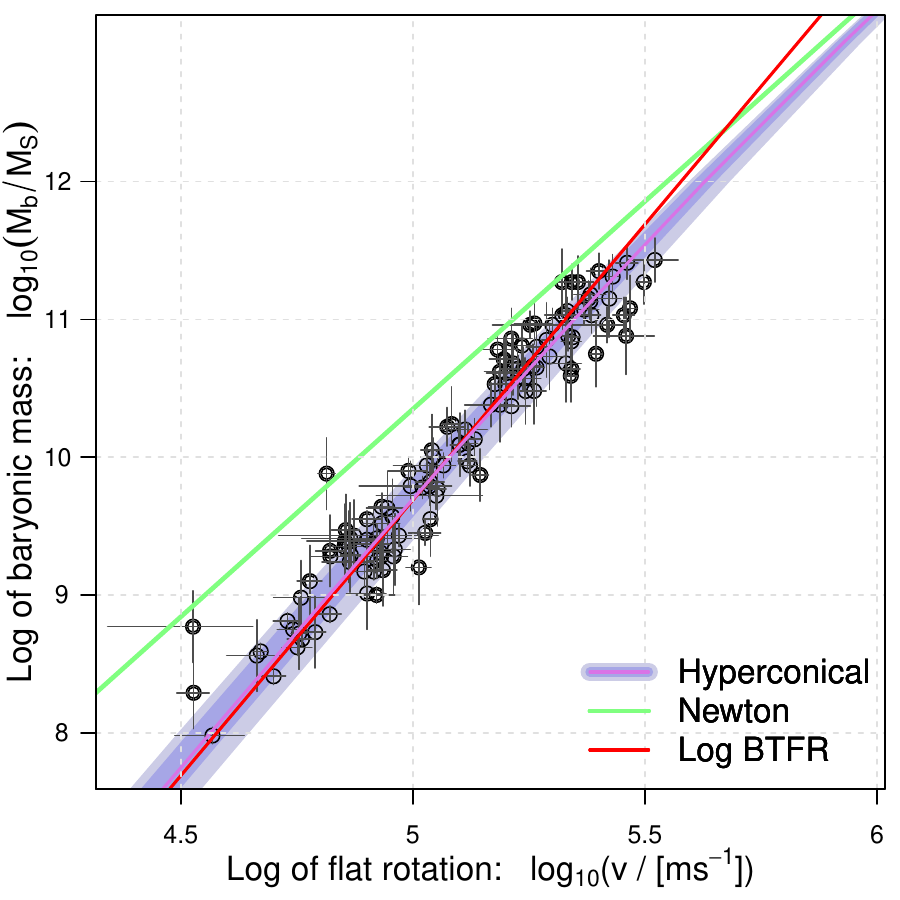}
	%\vspace*{8pt}
	\caption{MDAR/BTFR modeling. Fitting of the hyperconical model (purple line, Eq. \ref{eq:BTFR2} with $\gamma_M/\pi = 0.460 \pm 0.002$) to the baryonic Tully–Fisher relation $v(M_b)$ for 123 galaxies with `flat' velocity ($v$) and baryonic mass ($M_b$), expressed in terms of solar mass ($M_S$).  The Newton line represents the Keplerian orbits, while the ``Log BTFR'' line is the empirical law $M_b \propto v^\epsilon$ with $\epsilon \approx 4$. Finally, the used empirical relationship between the galactic radius $r_{M_b}$ and the baryonic mass $M_b$ is $r_{M_b} \propto M_b^{0.775\pm0.045}$. \label{fig:1}}
\end{center}
\end{figure}

MDAR data were obtained from the SPARC database, consisting of 153 late-type galaxies (spirals and irregulars) with Spitzer photometry and high-quality HI+Halpha rotation curves \cite{McGaugh2007, Lelli2016b, Lelli2019}. Similar to the cited studies, and compared to the original set of 175 galaxies of SPARC, 22 objects were excluded due to low inclinations and low-quality rotation curves. 

To represent the `flat velocity' of the rotation curves used in the MDAR/BTFR modeling, this paper used the same criterion as in Lelli et al. \cite{Lelli2019}, from which only 123 objects are used. The remaining 30 galaxies do not satisfy the 5\% flatness criteria, since their rotation curves are either rising or declining in the outer parts. This criterion is important to explore the sensitivity of modified gravity approaches for modeling dark-matter-like behaviors out to larger radii, where a flat part is reached in the rotation curve.

Individually for each galaxy, the parameter $\gamma_M$ has a large range of values with quartiles of about $\gamma_M/\pi = 0.457^{+0.013}_{-0.021}$, and the best fit (Eq. \ref{eq:discrepancy} or equivalently Eq. \ref{eq:BTFR}) to the data set is found for a constant $\gamma_M/\pi = 0.460 \pm 0.002$ (Fig. \ref{fig:1}).  To explain the mass discrepancy, this $\gamma_M$-hypercone-based model obtained an Adjusted R-squared of $R^2 = 0.92$, slightly greater than the simple logarithmic model of BTFR ($R^2 = 0.91$). The relative mean absolute error, $RMAE:=\mathrm{mean}(|modeled-observed|/observed)$, is also better for the $\gamma_M$-hypercone, obtaining $RMAE=0.108$ versus $RMAE =0.115$, estimated when the simple empirical BTFR logarithm is used. That is almost $\sim 1\%$ improvement with only a free parameter. 

%Mass models used in McGaugh et al. 2007, ApJ, 659, 149 ###### 
%#http://astroweb.case.edu/ssm/data/
%#https://iopscience.iop.org/article/10.1086/511807

\begin{figure}[pb]
\begin{center}
 \advance\leftskip-2cm
  \advance\rightskip-1cm
     %\makebox[\textwidth]{\includegraphics[width=\paperwidth]{Figure2_galaxy_rotation.pdf}}
    \includegraphics[scale=1.16]{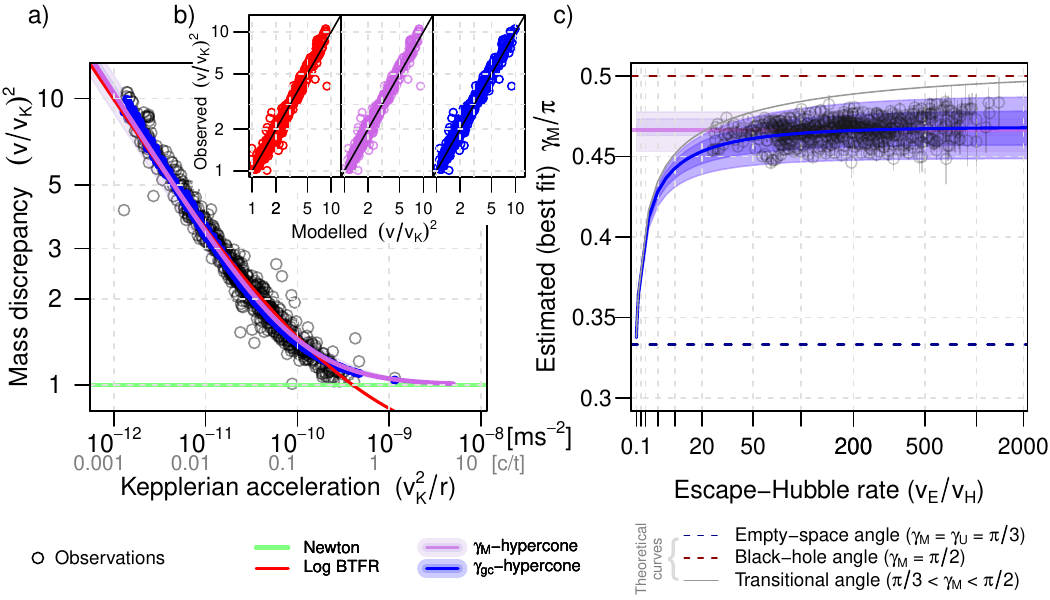}
	%\vspace*{8pt}
 
	\caption{Modeling of the rotation curve for 61 galaxies collected from the SPARC dataset: (a) Fit of Eq. \ref{eq:BTFR2} (purple line) and \ref{eq:BTFR3} (blue line) to the observed rotation curves; (b) Scatter plot of observed and modeled velocity according to the empirical ``Log BTFR'' (Eq. \ref{eq:BTFR0}), the $\gamma_M$-hypercone-based model (Eq. \ref{eq:BTFR2}) and the $\gamma_{gc}$-hypercone-based model (Eq. \ref{eq:BTFR3}) from \ref{eq:model_g1}); (c) Modeling of the fitted $\gamma_M$ according to a constant value of $\gamma_M/\pi \approx 0.466^{+0.011}_{-0.013}$ and $\gamma_{gc}$-hypercone, compared to the theoretical curves $\gamma_M = \gamma_{U}=\pi/3$ and $\gamma_M = \gamma_{gc}=\pi/2$. The `Newton' line represents the Keplerian orbital curves $v = v_K$, for which $M/M_b = 1 = v^2/v_K^2$.\label{fig:2}}
\end{center}
\end{figure}

For a set of 696 objects in 61 high-quality SPARC galaxies \cite{Lelli2019}, rotation curves were correctly simulated by the $\gamma_M$-hypercone-based model (Eq. \ref{eq:BTFR2}; Fig. \ref{fig:2}). The proposed model obtained a relative mean absolute error of $RMAE = 0.070$ and an R-squared of $R^2 = 0.955$, which are slightly better ($\sim 1\%$) than the empirical fitting ($RMAE = 0.080$ and $R^2 = 0.953$).

If individual dynamics of the 696 objects is considered, the range of values of $\{\gamma_M/\pi\} \sim (0.44, 0.48)$ is adequately modeled by the $\gamma_{gc}$-hypercone-based approach (Eqs. \ref{eq:model_g1}-\ref{eq:BTFR3}), and the observational constraint yields a constant $\gamma_{gc}/\pi = 0.468\pm 0.003$. In this case, the R-squared is $R^2 = 0.957$, but the relative error is the same as in the  $\gamma_M$-hypercone-based model ($RMAE = 0.070$).

The results showed that the hypercone-based galaxy curves have a transitional behavior between the deep MOND flat curves and the purely Newtonian-Keplerian regime. The $\gamma_M$-hyperconical approach is enough to adequately describe the transition between dynamics, but conceptually, the $\gamma_{gc}$-hypercone is more adjusted to limit cases of empty spaces and the black hole's environment. However, the difference between the fitted value of $\gamma_{gc}/\pi = 0.468\pm 0.003$ and the theoretical value of $\gamma_{gc}/\pi = \frac{1}{2}$ is still unexplained by the current state of the model. To confirm this deviation, further work could analyze the range of $\gamma_{gc}$ in black hole dynamics.

%http://doi.org/10.1111/j.1365-2966.2007.11433.x
%https://academic.oup.com/mnras/article/376/1/338/973651

\subsection{Empty space and solar system scales}

According to Eq. \ref{eq:Pioneer0}, our model predicts that empty spacetime experiences an acceleration of $a = c/(\gamma_0 t) \approx c/(2t) \approx 3.5\cdot 10^{-10}$ ms${}^{-2}$. This value is very close to the expected value by hypothetical Unruh-like radiation, with acceleration \begin{equation}
    a \sim \frac{\frac{4}{5}\pi^2 c^2}{\lambda} \sim  \frac{\pi^2 c}{10 t} \sim \frac{c}{t}\,,
\end{equation} where $\lambda = 4\Theta  = 8ct$ is taken following the Casimir-like effect proposed by \cite{McCulloch2007}, obtaining $6.9 \pm 3.5 \cdot 10^{-10} m s^{-2}$ (twice as much as our model).

However, perturbations of (vacuum) energy density cause a local curvature larger than the empty spacetime, reducing the cosmic acceleration. Taking into account the Eq. \ref{eq:model_g1} for the solar system, a projective angle of $\gamma_M \sim 0.48 \pi$ is predicted at a heliocentric distance of 1 UA and 30 UA, respectively, which corresponds to an inappreciable $a_p \sim 10^{-11}$ m/s${}^2$. 
 
This value is smaller than that observed in the Pioneer 10 and 11 spacecrafts. Radiometric tracking data collected from the Pioneer spacecrafts indicated the presence of an anomalous Doppler frequency drift interpreted as a constant acceleration of $(8.74 \pm 1.33) \cdot 10^{-10} ms^{-2}$ towards the Sun \cite{Turyshev2005}. This anomaly was explained several years later. The model based on thermal radiation pressure (TRP), and the corresponding thermal recoil force (TRF) can explain up to $8 \cdot 10^{-10}$ ms${}^{-2}$. Therefore, no statistically significant anomalous acceleration remains in the data, and it is up to 10\% with respect to the original anomaly in the Doppler effect \cite{RieversLammerzahl2011}. Nevertheless, the 10\% of uncertainty is of the same scale as predicted from Eq. \ref{eq:model_g0}.

Specifically, TRF and Doppler data can be modeled using an exponential decay model in the form $a = a_0 2^{-(t-t_0)/T}$ \cite{Turyshev2011, Turyshev2012}. Setting $t_0 =$ January 1, 1980, Turyshev et al.  \cite{Turyshev2011} estimated the parameters $T_{Dopp} = (28.8 \pm 2.0)$ yr, $a_{0, Dopp} =
(10.1 \pm 1.0) \cdot 10^{-10}$ m/s${}^2$ fitted to the Doppler data, while the ones for TRF are $T_{TRF} = (36.9 \pm 6.7)$ yr and
$a_{0,TRF} = (7.4 \pm 2.5) \cdot 10^{-10}$ m/s${}^2$. Again, the same uncertainty scale is found.

%the calculated thermal recoil force can be modeled, with an RMS error of 0.1×10−10 m/s2

%At the most, one can assumes that all the velocity components of Pioneer 10 and 11 are orthogonal to the Hubble expansion.

\subsection{Physical interpretation}

A conceptual parallel is possible to draw between the apparent acceleration of the expanding universe and the fictitious Coriolis acceleration of a non-inertial system like the Earth. The first one is produced when a reference frame changes from the ambient space (with an extrinsic viewpoint) to the tangent Minkowskian spaces, defined under the (intrinsic) metric of an observer. The second one is the well-known consequence of changing the reference from approximately inertial frames (e.g., heliocentric coordinates) to the rotating system of the Earth or the observers that live on its surface.

The features of the projected hyperconical (background) metric contribute to the large-scale dynamics of galaxies as the choice of reference frames is the key in non-inertial dynamics. For instance, the Coriolis force experienced by deviated trajectories on the Earth corresponds to the deviated redshifts experienced by Type Ia supernovae stars as a function of the luminosity distance. Similarly, the geostrophic wind defined in large-scale (extratropical) systems is mostly due to the Coriolis acceleration, as the `flat' speed curve of the spiral galaxies could be due to the apparent acceleration $c/t$ raised from the hypercone-based projection. The black-hole rotation curve would be the analog cyclostrophic wind of tropical cyclones, which are minimally linked to the Coriolis force compared to the extratropical systems \cite{John2006, Innes2022}. 

The cyclostrophic flow is the main contribution to the total wind when the centripetal acceleration (from gravity-related pressure gradient) is greater than the Coriolis acceleration, so it is observed in cases of very curved systems (such as a tornado or hurricane). Therefore, an open question on the apparent acceleration is its possible evidences at smaller cosmological scales. Specifically, possible anomalous acceleration in the solar system has been explored during the last decades, but no statistically significant differences were found with respect to the TRP/TRF-related anomalies observed in Pioneer 10 and 11 spacecrafts.

Moreover, the hyperconical background metric needs to be further studied since strong links are expected between the extrinsic/intrinsic hyperconical model and the $R_h=ct$ universe of Melia \cite{Melia2012, Melia2013}. For example, the cosmic timeline of the flat $R_h=ct$ model is exactly the same as in the linear expansion (extrinsic perspective) of the hypercone-based model \cite{Melia2023}, while the curvature is locally and globally compatible under the extrinsic and intrinsic viewpoint, respectively. The fictitious acceleration of the projected hyperconical manifold implies that the values of dark quantities are constant like the `zero active mass' of the Melia's universe \cite{Melia2017}. In fact, the linear expansion (e.g. $R_h=ct$ model) better predicts the formation of high-redshift galaxies than the standard $\Lambda$CDM \cite{Melia2023}, which is statistically incompatible with modeling the known as `impossible early galaxies' \cite{Yennapureddy2018, Ferreira2022, Gupta2023}.

The possible new paradigm seems to be also reinforced by recent observations of dwarf galaxies, collected from the Fornax Deep Survey catalog, which showed that their deformation and lack of low surface brightness dwarfs towards its center are incompatible with $\Lambda$CDM expectations but well consistent with MOND \cite{Kroupa2012, Kroupa2022, Asencio2022}. Tidal dwarf galaxies are formed from baryonic material rapidly ejected during the tidal interaction of two galaxies and, according to the dark matter hypothesis, this process is too fast to allow for efficient capture of dark matter and hence they must not present mass discrepancy. However, Kroupa \cite{Kroupa2012} found that three objects appear to have mass discrepancies that are in close agreement with the MOND prediction. Moreover, Kroupa et al. \cite{Kroupa2022} claim that asymmetry in the population of leading-trailing tidal tails and observed lifetime of open star clusters are inconsistent with Newtonian dynamics but consistent with MOND. On the other hand, dark-matter-deficient relic galaxies have velocity curves outside the MOND-regime domain and, therefore, they do not exclude modified gravity theories \cite{Comeron2023}. Following the meteorological analogy, these cases are like cyclostrophic storms with negligible interaction with the Coriolis (i.e., MOND-like) acceleration.

According to our results, paths of both the light and the gravitational interaction in galaxies would be drawn on the (intrinsic) manifold as the atmospheric winds represent flows and interactions of air masses on the Earth's surface. Nonetheless, the cosmic timeline of both systems is defined by the extrinsic reference frames. Therefore, as a candidate for relativistic MOND theory, the present work opens new frontiers and challenges in astrophysics and cosmology.

\section{Conclusions}

As dark energy can be interpreted as a geometrical consequence of the distorted stereographic projection (according to \cite{MCS2020, MCS2023}), this work shows that dark matter can be also modelled by the same projection, and it is enough using the first-order perturbation approach. To build the proposed model, from hyperconical universes embedded in 5D Minkowskian spacetime, two additional ingredients are required: (1) massive objects are defined as perturbations of the vacuum energy density, and (2) centrifugal force has an extra orthogonal contribution due to the curvature radius $t$ of the Universe. This corresponds to a time-like component in the total centrifugal force, whose squared modulus is proportional to $v^4$, and the term $r/(\gamma_0ct)$ of the distorted projection contributed to it. Thus, the distorted stereographic projection contributes to deep MOND behaviour with the acceleration $c/t$ divided into the projected angle $\gamma_0$. 

Using SPARC data, the proposed fictitious force explains the mass discrepancy ($M/M_b$) slightly better ($\sim 1\%$) than the BTFR empirical fitting ($RMAE =0.115$ and $R^2 = 0.91$  versus $RMAE=0.108$ and $R^2 = 0.92$, respectively) and also better for galaxy rotation curves ($RMAE=0.080$ and $R^2 = 0.953$  versus $RMAE=0.070$ and $R^2 = 0.957$, respectively), even considering a unique parameter. Furthermore, the new model explains the transition between the Newtonian and the cosmological scales in a natural way. Therefore, the proposed theoretical framework is a candidate for the relativistic formulation of the MOND behavior.

It is remarkable that our goal is achieved by employing two key points (the mass of perturbations and the centrifugal force) but with a minimal change in the Lagrangian density of the GR. The proposal is to adapt the background metric to the hyperconical universe (Eq. \ref{eq:exact_line}), since its curvature scalar is equal to the vacuum energy density. Specifically, under the first-order perturbation approach, our proposal of linearized modified gravity is given by $h_{\mu\nu} :\approx g_{\mu\nu} - g_{\mu\nu}^{back}$ where $g_{\mu\nu}$ is the perturbed hyperconical metric (see, e.g., Eq. \ref{eq:Schwarzschild}) while $g_{\mu\nu}^{back}$ is the background metric (Eq. \ref{eq:exact_line}), which replaces the usual flat 4D Minkowskian metric $\eta = \mathrm{diag}(1,-1,-1,-1)$. This linearization is equivalent to say that matter content does not affect the shape of the universe but locally perturbs the metric. That is, the Lagrangian density of matter (i.e. $\mathcal{L}_M = \rho_M$) is linked to the difference $\Delta R = R-R_u$ between the total Ricci scalar and the universe's Ricci scalar $R_u \approx -6/t^2$, as proposed in \cite{MCS2020}. As an example, the present paper developed and used a procedure (\textbf{Appendix B}) that recovers a Schwarzschild-like vacuum solution as a limit case of the hyperconical metric perturbed by a central mass. On the other hand, the distorted stereographic projection is a key for recovering deep MOND regimes under the hyperconical model. In particular, the resulting new coordinates experience a fictitious acceleration like that modeled by MOND theories.

Further work could analyze possible observations related to the predicted anomalous accelerations in the solar system of about $10^{-11}$m/s${}^2$, according to the theoretical framework developed in this paper. Particularly, it is possible that residual unexplained accelerations (about 10\%) of the Pioneer 10 and 11 spacecrafts (or of other objects) are due to the time-like contribution of centrifugal force in the hyperconical universe. Thus, the present work is a candidate for a modified gravity model which is more precise than GR for galaxy rotation calculations, and it would be useful to better estimate the Newtonian constant of gravity. Finally, connections between the proposed model and Melia's $R_h=ct$ universe \cite{Melia2017} could be deeply explored, especially concerning the role of `zero active mass' in the theory developed. 

\section*{Acknowledgments}

The author is very grateful to Prof. R. Campoamor-Stursberg, Prof. P. Kroupa and the anonymous referees for their helpful comments. It is very fair to dedicate distinguished recognition to the first anonymous referee due to all the excellently suggested amendments.

\section*{Appendix A. Lagrangian density with perturbed vacuum energy}
%https://arxiv.org/pdf/2011.04175.pdf

\noindent %Despite natural units ($c=1=h$) are preferred in our work, this section explicitly uses the constants $c$ and $h$ to facilitate the interpretation of the results. 

This appendix aims to define the Lagrangian for the modified gravity theory considered in this paper.  

Assuming that GR is valid at local scales, total Lagrangian density is obtained by modifying the standard Einstein-Hilbert term \cite{MCS2020},
\begin{eqnarray} \label{eq:space.lagrangian}
\mathcal{L} = \frac{1}{16\pi\mathrm{G}} \Delta R +
\mathcal{L}_M  \; \approx \; \frac{1}{16\pi\mathrm{G}}\left(R + \frac{6}{t^2} \right) -
\rho_M \;  \approx \; \frac{c^2}{16\pi\mathrm{G}}R - \Delta \rho\,,
\end{eqnarray}
where $\Delta R := R - R_u$ is the curvature perturbation with respect to the local limit of the Ricci curvature of the (empty) hyperconical universe, $R_u \approx -{6}/{t^2}$, while $\mathcal{L}_M = -\rho_M$ is the Lagrangian of mass-energy, and $\Delta \rho = \rho_M - \rho_0$ is the density perturbation compared to the ``vacuum energy'' $\rho_0 =
3/(8\pi\mathrm{G}t^2)$ and its mass-related event radius is $r_{\mathcal{M}} := 2\mathrm{G} \mathcal{M} =  2\mathrm{G} \rho_0 \, \frac{4}{3} \pi t^3  = t$. Moreover, the squared escape velocity linked to $\rho_0$ at $r$ is $v_E^2(\rho_0) =  2\mathrm{G} \rho_0 \, \frac{4}{3} \pi r^3 = r^2/t^2 = v_H^2$. Therefore, total density $\rho_M$ leads to a total squared escape velocity $v_E^2(\rho_M)$ as follows\begin{equation}
v_E^2(\rho_M) = 2\mathrm{G} \rho_M \, \frac{4}{3} \pi r^3 = 2\mathrm{G} \left(\rho_0 + \Delta\rho  \right)\, \frac{4}{3} \pi r^3 = \frac{r^2}{t^2} + \frac{2\mathrm{G}M}{r} = v_E^2(\rho_0) + v_E^2(\Delta\rho)\,, \;\;
\end{equation} where it has been used that $M := \Delta\rho \, \frac{4}{3} \pi r^3\;\propto\; \rho_0 t^3 \;\propto\; t$.

Now, let $\theta_M := M/\mathcal{M} \ll 1$ be a (small) constant fraction of energy corresponding to the perturbation $\Delta \rho$, and $r_M := 2\mathrm{G}M = \theta_M t$ be the mass-related event radius. Thus,\begin{equation}
\frac{2\mathrm{G}M}{r} = \frac{\theta_M t}{r'\frac{t}{t_0}} = \frac{\theta_M t_0}{r'}=: \frac{2 \mathrm{G}M_0}{r'}\,.
\end{equation} Therefore, the quotient $M/r = M_0/r'$ is as comoving as $r/t = r'/t_0$.

Moreover, background metric of the universe leads to $R_{00}^u = 0$ and $R_{ij}^u = \frac{1}{3} {R}_u \, g_{ij}$. Since $R_u = -\frac{6}{t^2}$, Einstein field equations become locally converted to\begin{equation}
    \left\lbrace
\begin{array}{l}
\kappa T_{00} \approx R_{00} - \frac{1}{2} R g_{00}, \\[2ex]
\kappa T_{ij} \approx \Delta R_{ij} - \frac{1}{2} \Delta R g_{ij} \approx R_{ij} - \frac{1}{2} R g_{ij} - \frac{1}{t^2}.
\end{array}
\right.
\label{eq:modified_EFE1}
\end{equation} where $\kappa = 8\pi\mathrm{G}$ and $T_{\mu\nu}$ are the stress-energy tensor components. Notice that the new term $-\frac{1}{t^2}$ behaviours like a dynamical dark pressure (varying as $a^{-2}$).

\section*{Appendix B. Schwarzschild limit}

The present appendix develops the hyperconical metric with a perturbation that recovers a limit solution similar to the Schwarzschild vacuum solution. 

Let $(t, \vec{r}, u) := (t, x, y, z, u) \in \mathbb{R}_{\eta}^{1,4}$ be Cartesian coordinates, including an extra spatial dimension $u$ in the five-dimensional Minkowski plane. Specifically, $u :=  t \cos\gamma  - t$ is chosen to mix space and time with gravity of central mass $M$ observed at a distance $\hat r$ such that $\sin^2 \gamma := \frac{r^2}{t^2} = \frac{r'^2}{t_0^2}  =: \frac{\hat r^2}{t_0^2} + \frac{2\mathrm{G}{M}}{\hat r}$ is defined for some small constant $t_0 \in \mathbb{R}$ and comoving distance $r' := \frac{t}{t_0} r$. Thus, the line element is \begin{eqnarray} \label{eq:forma_rapida0}  \nonumber
ds^2 &= & (dt, d\vec r, du)^2 \; = \; dt^2 - d\left(r \vec e_r \right){}^2  -du^2 \; = \; 
\\ \nonumber
& =& dt^2 - dr^2 - r^2d\Sigma^2 - d\left( t \cos \gamma - t\right) ^2 \,,
\end{eqnarray}where $\vec e_r := \vec r/r := \vec r'/r' =: \vec {\hat r} /\hat r$ is the unitary spatial vector. The extra spatial dimension $du$ contributes to: \begin{eqnarray}  \nonumber
du  &=& d\left( t \cos\gamma - t\right)  = dt\left( \cos\gamma - 1\right) - t\sin\gamma d\gamma
\\  \nonumber
du^2  &=& dt^2\left( \cos^2\gamma -1\right)^2 + t^2\sin^2\gamma\,  d\gamma^2 - 2t\left( \cos\gamma - 1\right)\sin\gamma\, d\gamma dt
\end{eqnarray}and adding the differential of the radial coordinate, \begin{eqnarray}   \nonumber
dr &=&  d\left(t\sin\gamma \right) = \sin\gamma dt + t\cos\gamma d\gamma
%%%  \\ \nonumber
%%% dr^2 &=& \sin^2\gamma dt^2 + t^2\cos^2\gamma d\gamma^2 + 2t\sin\gamma\cos\gamma dt d\gamma
\\ \nonumber
&\Rightarrow & du^2 + dr^2 =  dt^2\left( 2-2\cos\gamma\right) + t^2 d\gamma^2 + 2t \sin\gamma\, d\gamma dt
\end{eqnarray}Finally, total differential for $x^\mu = (t, r', \theta, \varphi) \in \mathbb{R}_{\hat g}^{1,4}$ is given by \begin{eqnarray} \nonumber
    ds^2 = dt^2 - d\vec{r}'^2 - du^2 = \hat g_{tt}dt^2 + \hat g_{r'r'}dr'^2 +  \hat g_{\theta\theta}d\theta^2 + \hat g_{\varphi\varphi}d\varphi^2 + 2\hat g_{r't}dr'dt
\end{eqnarray} Thus, using that $t_0 d\gamma  =  dr'/\cos\gamma$, \begin{eqnarray} \nonumber
   \hat  g_{tt} &=&  2\cos\gamma - 1 \;\; \approx   \;\; 1 - \frac{{\hat r}^2}{t_0^2} - \frac{2\mathrm{G}M}{{\hat r}}
     \\ \nonumber
\hat g_{r'r'}  &=&  - \frac{t^2}{t_0^2} \frac{1}{\cos^2\gamma} 
\; = \;
- \frac{t^2}{t_0^2} \left(1 - \frac{{\hat r}^2}{t_0^2} - \frac{\mathrm{2G}{M}}{{\hat r}} \right)^{-1} 
\approx  
- \frac{t^2}{t_0^2} \left(1 + \frac{{\hat r}^2}{t_0^2} + \frac{\mathrm{2G}{M}}{{\hat r}} \right)
\\ \nonumber
\hat g_{r't}  &=& \frac{t}{t_0}\tan\gamma \;\;  =\;\;  \frac{t}{t_0}\frac{r'}{t_0} \left(1 - \frac{{\hat r}^2}{t_0^2} - \frac{2\mathrm{G}{M}}{{\hat r}} \right)^{-1/2} \;  \approx \; \frac{t}{t_0}\frac{r'}{t_0} + O\left(\frac{{\hat r}^3}{2t^3}\right)
\\ \nonumber
\hat g_{\theta\theta} &=& - \frac{t^2}{t_0^2} r'^2
\\ \nonumber
\hat g_{\varphi\varphi} &=& - \frac{t^2}{t_0^2} r'^2 \sin^2\theta\,.
\end{eqnarray} Therefore, assuming linearized perturbations of the metric $\hat g_{\mu\nu} =  \hat g_{\mu\nu}^{back} + h_{\mu\nu}$ with $\hat g_{\mu\nu}^{back} := \hat g_{\mu\nu}|_{M=0}$, we can find a local approach to the Schwarzschild metric perturbation $h|_{Schw}$ as follows \begin{eqnarray} \label{eq:Schwarzschild_h}
       h_{\mu\nu}\big|_{Schw}  & :\approx  &  \hat g_{\mu\nu} - \hat g_{\mu\nu}\big|_{M  \to 0} 
   \\   \label{eq:Schwarzschild_g}
  ds^2\big|_{Schw}  & \approx  &  \left(1- \frac{2\mathrm{G}{M}}{\hat r} \right) dt^2
-  \frac{t^2}{t_{0}^2} \left[\left(1 + \frac{2\mathrm{G}{M}}{\hat r}\right) d{r'}^2 + {r'}^2d{\Sigma}^2 \right] \,.\;\;
\end{eqnarray} Notice that $r$ and $r'$ are coordinates linked to the mass $M$, in contrast to the observed radial distance $\hat r$. Particularly, $r' = \sqrt{{\hat r}^2 + t_0^2\frac{2\mathrm{G}{M}}{\hat r}}$ leads to\begin{equation}
    dr' = \frac{{\hat r} - t_0^2\frac{\mathrm{G}{M}}{{\hat r}^2}}{\sqrt{{\hat r}^2 + t_0^2\frac{2\mathrm{G}{M}}{\hat r}}}  d{\hat r}
\end{equation} where $0 < t \approx t_0 \ll 1$ need to be sufficiently small to obtain $dr' \approx d\hat r$ and, therefore, they are initialized with a different criteria that the age of the universe.

\section*{Appendix C. Distorted stereographic projection}
%https://arxiv.org/pdf/2011.04175.pdf

This appendix derives the distorted stereographic projection applied to the hyperconical universe, which assimilates the radial inhomogeneity as a fictitious acceleration (i.e. an effect of a projective angle). 

Let $\alpha$ and $\lambda$ be (both positive) `distortion parameter' and `scale factor', respectively, that transforms time $t \mapsto \hat{t} := t\lambda$ and distorts the comoving length as $r' \mapsto \hat r' := r'\lambda^{\alpha}$. According to local consistency in dynamical systems \cite{MCS2023}, the distorted parameter needs to be $\alpha=\frac{1}{2}$, and its best fit was $\alpha = 0.499\pm 0.013$ when Type Ia Supernovae data were used. Moreover, let $F_q : \mathbb{R} \to \mathbb{R}^5$ be a pencil with $F_q(\lambda) \in \mathbb{R}^5$ parameterized, such as $F_q(1) = q(t) = (t, \vec{r}, u)$ and $F_q(0) = \hat q(0)  := (0, \vec{0}, u_0)$, with $u_0 := - t_0$ and $t_0$ being the age of the Universe. A transformation, given by  $\vec{r}' = r'\vec{e}_{r} \; = \;  t_0 \sin  \gamma\, \vec{e}_{r} \; \mapsto \; \hat r'\vec{e}_{r} := t_0 \sin \hat \gamma\, \vec{e}_{r}$, is performed for the angles $\gamma \mapsto \hat \gamma$, preserving the direction $\vec{e}_{r}$, and lengths are transformed as follows:\begin{equation}
F_{q}^{(a)} =
\left\{
\begin{array}{l}
\hat{t} = t \lambda, \\[1ex]
\hat{r}' = r' \sqrt{\lambda}, \\[1ex]
\hat{u} = u_0 + (u - u_0) \lambda.
\end{array}
\right.
\end{equation} When the points are projected on the `observer hyperplane', $\hat u = t_0$, a solution for the distorted stereographic projection is given by some $\lambda = \lambda_s(t,\gamma)$:\begin{equation} \label{eq:lambda}
t_0 = - t_0 + \left( t\cos\gamma  +  t_0 \right)\lambda_s % \;\; \Rightarrow\;\; \lambda_s(t,\gamma) = \frac{2}{(1+\frac{t}{t_0}\cos\gamma)}
\end{equation} Therefore, the search for solutions for $\lambda$ in $r' \mapsto \hat r' := r'\sqrt{\lambda}$ is reduced to find the geometrical relation between time $t$ and the comoving coordinate $r'$ or $\gamma$. Applying the global relationship $t(\gamma)$ proposed in Monjo and Campoamor-Stursberg \cite{MCS2023}, it is found that $u(t,\gamma) = t\cos\gamma$ is given by:\begin{eqnarray} \nonumber
     \gamma = 0 &\;\Rightarrow\;\;&  t=t_0\lambda^{-1} 
     \;\;\Rightarrow\;\; u(t,0) =  u(t_0,0) \lambda^{-1}\;
     \\ \nonumber
     \gamma = \gamma_{\max} &\;\Rightarrow\;\;&\frac{ t_0 \lambda^{-1} - t(\gamma_{\max})}{t_0} \le 1 \;\;\Rightarrow\;\; \frac{ u(t_0,\gamma_{\max}) \lambda^{-1} -u(t,\gamma_{\max})}{u(t_0,\gamma_{\max})} \le 1 
     \\ \label{eq:tgamma}
     \gamma \in [0, \gamma_{\max}) &\;\Rightarrow\;&\frac{ u(t_0,\gamma) \lambda^{-1} -u(t,\gamma)}{u(t_0,\gamma_{\max})} \in [0,1) \,,
    %However, in general $\lambda_s \neq t_0/t$, Thus, $t_U \equiv -\frac{2}{3}t_0$ corresponds to the angle  $\gamma=\gamma_U$ with $\lambda=3$.
\end{eqnarray} where the maximum angle $\gamma_{\max} \sim \gamma_U = \pi/3$ is chosen to be the domain limit for an empty space, and $t_0 \lambda^{-1} - t \le t_0$ is the maximum difference allowed.  Linearly with $\gamma$, the $t(\gamma)$ relationship of Eq.  \ref{eq:tgamma} is approximately given by\begin{equation} \label{eq:global}
 (t_0\lambda^{-1} - t )\cos \gamma \approx t_0\cos\gamma_U  \,\frac{\gamma}{\gamma_U} \approx: t_0  \,\frac{\gamma}{\gamma_0}\,.
 \end{equation} with projected angle $\gamma_0 := \frac{\gamma_U}{\cos\gamma_U} = \frac{2}{3}\pi \sim 2$ for empty spaces. Isolating $t$ from the above expression, Eq. \ref{eq:lambda} is now: \begin{equation} \label{eq:lambda_global}
t_0  \approx - t_0  + \left( t_0\left(\frac{1}{\lambda_s} - \frac{\gamma }{\gamma_0 \cos\gamma}\right) \cos\gamma  +  t_0 \right)\lambda_s\;\;\; \Rightarrow\;\;
\lambda_s  \approx  \frac{2 - \cos\gamma}{1-\frac{\gamma}{\gamma_0}} \sim \frac{1}{1-\frac{\gamma}{\gamma_0}}  \,.
\end{equation}Similar results to Eq. \ref{eq:lambda_global} are found if the local projection is used instead of the global approach, with a theoretical value of $\gamma_0 = 2$  and an observational constraint of $\gamma_0 = 1.6^{+0.4}_{-0.3}$ fitted to Type Ia Supernovae redshift \cite{MCS2023}.

\end{document}